# Using social psychological interventions to make physics classes equitable and inclusive

Chandralekha Singh

Department of Physics and Astronomy, University of Pittsburgh, Pittsburgh, PA 15260

**Abstract.** This article summarizes why physics instructors should use social psychological interventions to make physics classes equitable and inclusive.

When students struggle to solve challenging physics problems, they can respond in two distinct ways. One type of (negative) reaction is to question whether they have what is needed to excel in physics. A different (positive) reaction is to enjoy the struggle because it means the student is tackling new physics and learning. The negative reaction is a manifestation of fixed mindset (i.e., believing that intelligence is immutable and struggling reflects lack of intelligence), whereas the positive reaction emanates from a growth mindset (the fact that your brain's capabilities can grow with deliberate effort and you can become an expert in a field by working hard and smart) [1]. Unfortunately, due to societal stereotypes [2], women and ethnic and racial minority (ERM) students who are severely underrepresented in physics [3-27] are more likely than the majority students to fall prey to the fixed mindset trap and view struggle with challenging physics problems in a negative light. This is not surprising because compared to any other STEM field, the societal stereotypes are the strongest in physics, a field which has historically been associated with brilliant men. These stereotypes contribute to a lower sense of belonging for women and ERM students in physics learning environments [3-27].

To help players excel in any game, e.g., chess, coaches must ensure that the players have both good defense and offense. Helping students learn physics well, like helping players do well in a game, requires instructors to ensure that they equip all students with both good defense and offense. In particular, instructors should consider strengthening students' defenses by creating learning environments where all students have a high sense of belonging, promoting and emphasizing growth mindset, and ensuring that all students have high self-efficacy to excel in physics. Only if students have strong defenses pertaining to physics learning can they effectively engage with the offense, e.g., by tackling challenging problems and developing physics problem solving, reasoning and meta-cognitive skills.

Students with weak defenses are unlikely to undertake the risk of struggling with challenging physics problems. Without strong defenses, tackling challenging physics problems can collapse a "student's wavefunction" into a state in which the outcome is negative and the student contemplates: "I am struggling because I do not have what it takes to do well in physics. What is the point of even trying?". These kinds of negative thoughts can lead to a lack of engagement with effective approaches to learning physics and can increase students' anxiety during test taking so that some of the limited precious cognitive resources during problem solving are occupied by the anxiety pertaining to solving challenging problems. Unless instructors help all students develop adequate defenses, students with lower defenses can go in a detrimental feedback loop in which negative thoughts about struggling lead to increased anxiety, procrastination, and disengagement from effective learning approaches including taking advantage of the available resources for learning. The result is deteriorated performance which can then lead to further negative thoughts and anxiety. Due to societal stereotypes and biases that people are bombarded with from a young age [2], women and ERM students are less likely than majority students to have strong defenses when they enter physics classes. Therefore, if the instructor does not make a concerted effort to bolster student defenses and inoculate students against stereotype threats (i.e., fear of confirming a negative stereotype about one's group), the situation is more likely to hurt women and ERM students [2].

Fortunately, instructors and advisors have the power to empower students and impact their defenses positively by creating an inclusive and equitable learning environment in which all students have a high sense of belonging, where students are not afraid to struggle and fail, and where students use their failures as a stepping stone to learning [28-30]. Although physics instructors have traditionally not considered it to be their responsibility to serve as coaches for their students and help boost them along both the defense and offense dimensions pertaining to learning physics, these issues are central for equity and inclusion in physics. Moreover, short classroom activities that take less than a class period at the beginning of the course can go a long way in improving students' sense of belonging and intelligence mindset, particularly for those who need it the most and in creating an equitable and inclusive physics learning environment [28-30].

We implemented a short intervention that shows great promise [28] that only requires half of a recitation class period at the beginning of the semester. Our intervention was conducted in a required introductory calculus-based physics course, which is taken by physical science and engineering majors typically in their first year first semester in college. Two female physics graduate students were trained to facilitate the half-hour activity at the beginning of the semester in half of the recitations that were randomly selected. The facilitators introduced it as an activity that would help the physics department understand student concerns and how to foster better learning environments. Students in the recitation classes in which the activity took place were handed a piece of paper and asked to write about their concerns about being in the physics course. Then they were shown some quotations from both male and female students from previous years who did very well in physics who also had similar concerns. The quotes emphasized the importance of working hard and working smart, learning from one's mistakes and taking advantage of all of the learning resources because that is the way to perform well in physics. Then students were asked to get together in small groups to discuss what they wrote; generally, they learned that other students in their classes had similar worries. Finally, there was a general class discussion summarizing what the different groups discussed, with explicit emphasis of the fact that adversity is common in college physics courses but it is temporary. The facilitators re-emphasized that students should embrace challenging physics problems, use their failures as stepping stones to learning and work hard and work smart to succeed. Using the principle of "saying is believing" [28-29], in the next recitation class, students were asked to write a short letter telling a future student about strategies for excelling in their physics class.

What is heartening is that this short intervention closed the gender gap in performance compared to the comparison group involving the recitation sections in which this short intervention did not take place [28]. A student's sense of belonging, self-efficacy and intelligence mindset are strongly intertwined with cognitive engagement and learning [1,2]. Just after winning the US Open in 2019, Naomi Osaka proclaimed, "Fall on my face 18 million times and I'm gonna get up 18 million times. Just wanted to say I'm probably gonna fall down a couple dozen times in the future but hey, the kid is resilient." Without improving students' defenses, it is impossible for them to use their cognitive resources appropriately and excel in physics. Physics instructors should consider activities similar to the one we implemented [28] that strive to create classrooms that are inclusive and equitable and give all students an opportunity to develop a solid grasp of physics. Last but not least, it is important to remember that the authenticity and credibility of the facilitator of the activity (e.g., instructor or teaching assistant) is extremely important for students to trust the message underlying the activity and benefit from it.

**Acknowledgement**

I am very grateful to Dr. Kevin Binning for his help in implementing the intervention in physics courses.